\documentclass[a4paper]{llncs}

\usepackage{textpos}
\usepackage{blindtext}
\usepackage{placeins}
\usepackage[dvipsnames]{xcolor}
\usepackage{amsthm}
\usepackage[francais,english]{babel}
\usepackage{framed}
\usepackage{url}
\usepackage[toc,page]{appendix}
\usepackage{multirow,array}
\usepackage{soul}

\usepackage[utf8]{inputenc}

\usepackage{amsfonts,amstext,amsmath,amssymb}

\usepackage{graphicx}
\usepackage{wrapfig}

\usepackage{enumerate}

\usepackage{listings}
\usepackage{tgcursor}
\usepackage{multirow}

\usepackage{color}
\usepackage{cancel}
\usepackage{caption}
\usepackage{float}
\usepackage{graphics}
\usepackage{fancyhdr}
\usepackage{subfig}

\begin{document}

\title{On measuring performances of C-SPARQL and CQELS}
%\title{Experiment of Performance Evaluation on C-SPARQL and CQELS}

\author{Xiangnan Ren$^{1,2,3}$, Houda Khrouf$^{1}$, Zakia Kazi-Aoul$^{2}$, Yousra Chabchoub$^{2}$ \and Olivier Curé$^{3}$ }
\institute{	
 $^{1}$ ATOS - 80 Quai Voltaire, 95870 Bezons, France\\
 \email{\{xiang-nan.ren, houda.khrouf\}@atos.net}\\
 $^{2}$ ISEP - LISITE, 75006 Paris, France\\
 \email{\{zakia.kazi, yousra.chabchoub\}@isep.fr}\\
 $^{3}$ UPEM LIGM - UMR CNRS 8049, 77454 Marne-la-Vallée, France  \\
  \email{olivier.cure@u-pem.fr}
 }
			
\maketitle
\begin{abstract} 
To cope with the massive growth of semantic data streams, several RDF Stream Processing (RSP) engines have been implemented. The efficiency of their throughput, latency and memory consumption can be evaluated using available benchmarks such as LSBench and CityBench. Nevertheless, these benchmarks lack an in-depth performance evaluation as some measurement metrics have not been considered.
%They also do not clearly distinguish the mechanisms such as data-driven and time-driven employed by an RSP engine. 
%In this work, we extend existing benchmarks with novel performance criteria taking into account the targeted RSP mechanism.
The main goal of this paper is to analyze the performance of two popular RSP engines, namely C-SPARQL and CQELS, when varying a set of performance metrics. More precisely, we evaluate the impact of stream rate, number of streams and window size on execution time as well as on memory consumption. %The results lead us to deeply understand these RSP engines and will help us to design our own system.

%\textbf{Keywords:} RDF Stream processing, continuous queries	, Semantic Web
\end{abstract}

\section{Introduction} \label{sec:introduction}
With the emergence of Big data's velocity aspect, new platforms are needed to efficiently handle data streams. In the context of the Semantic Web, a dedicated W3C community\footnote{\url{https://www.w3.org/community/rsp/}} extended standard SPARQL queries with the ability to continuously query unbounded RDF streams. This is a key component of RDF (Resource Description Framework) Stream Processing, henceforth denoted as RSP. The development of RSP engines integrates some streaming features such as windowing operations and periodical execution. Examples of popular RSP engines are C-SPARQL~\cite{CSPARQL} and CQELS~\cite{CQELS}. Each engine has its specific architecture, query language, execution mechanism and operational semantics. In order to determine which RSP engine to adopt for a particular application, it is primordial to conduct an in-depth and complete performance analysis.

Since 2012, benchmarks and comparative research surveys of RSP engines have been conducted. Examples of RSP benchmarks are SRBench~\cite{SRBench}, CSRBench~\cite{CSRBench}, LSBench~\cite{LSBench} and CityBench~\cite{CityBench}. While SRBench and CSRBench have studied query functionalities and output correctness, LSBench and CityBench go a step further by tackling performance criteria. However, current benchmarks do not distinguish between the different mechanisms, namely time-driven and data-driven described in \cite{DBLP:journals/pvldb/BotanDDHMT10} and employed by existing RSP engines. Moreover, some performance criteria have not been considered in the evaluation plan of these benchmarks. Thus, we conduct some experiments to have a deeper comprehensive view on current RSP systems.
% Thus, we conduct a redesigned experiment to have a comprehensive view on current RSP systems. 
%Generally, we consider that C-SPARQL and CQELS belong to two different categories depending on report policy: time and data driven.
We do not propose a new benchmark, but we propose an evaluation of some performance criteria. More precisely, we evaluate the impact of stream rate, window size, number of streams, number of triples and static data size, on query execution time and memory consumption. This evaluation has been conducted on the two popular RSP engines: C-SPARQL and CQELS.

%Thus, we propose an in-depth evaluation plan, as an extension of existing benchmarks, in which we consider the different RSP mechanisms and we target several performance criteria, i.e., impacts of stream rate, window size, number of streams, number of triples, and static data size on query execution time and memory consumption which are all further detailed in the remaining of this paper.

This paper is organized as follows. Section \ref{sec:related-work} provides an overview of existing RSP engines and benchmarks. In Section \ref{sec:evaluation-plan}, we describe our evaluation plan and novel performance criteria. Section \ref{sec:experiment} presents the results of our experiments, and we discuss them in Section \ref{sec:result-discussion}. We conclude and outline future work in Section \ref{sec:conclusion}.

\section{Related Work} \label{sec:related-work}
In this section, we first present the two popular RSP engines, C-SPARQL and CQELS, and then we describe existing RSP benchmarks.

\subsection{RSP engines}
C-SPARQL~\cite{CSPARQL} and CQELS~\cite{CQELS} represent two mature and mostly used RSP engines. Each engine proposes its own continuous query language extensions to query time-annotated triples and employs a specific RSP mechanism. We precisely distinguish two kinds of RSP mechanisms: time-driven and data-driven. The time-driven mechanism periodically executes SPARQL queries within a logical window (time-based) or physical window (triple-based). Whereas, the data-driven mechanism executes SPARQL queries immediately after the arrival of new data streams. In the following, we present the main features supported by each aforementioned engine.

\textbf{C-SPARQL} supports time-driven query execution and extends the standard SPARQL query language with keywords such as \texttt{RANGE} and \texttt{STEP}. The \texttt{RANGE} keyword defines the time-based window (e.g., \texttt{RANGE 5m} means a window of 5 minutes), and the \texttt{STEP} keyword indicates the frequency at which the query should be executed. Standard SPARQL 1.1 operators can be used over the data within the window such as aggregation, ordering and comparison. C-SPARQL streams out the whole output at each query execution, which refers to Rstream operator among the different streaming operators~\cite{Arasu:BW03} (e.g., Rstream, Istream, Dstream). 

%By consequence,  the output can be more verbose, as the same solutions can be present at different execution times (i.e., the same solutions can be present at \texttt{STEP 1} and \texttt{STEP 2}).

\textbf{CQELS} is developed in a native and adaptive way proposing a pre-processor and an optimizer to improve performance~\cite{CQELS}. It supports data-driven query execution following the \emph{content-change policy}, in which queries are triggered immediately at the arrival of new statements in the window. Even if the \texttt{SLIDE} keyword is supported in CQELS syntax (like \texttt{STEP} keyword in C-SPARQL), it does not have any effect on the engine behavior. The frequency execution depends on the arrival of new data in the stream. CQELS compares the current output with the previous one, and streams out only the new results, which refers to Istream~\cite{Arasu:BW03} operator in terms of streaming operators.

% \begin{table}
% \renewcommand{\arraystretch}{1.4}
% \centering{
% \begin{tabular*}{8cm}[10pt]{c @{\extracolsep{\fill}} ccc}
% \hline 
% \textbf{RSP} & \textbf{Mechanism} & \textbf{R2S operator}  \\ 
% \hline
% \textbf{C-SPARQL} & Time-driven & RStream \\ 
% \hline
% \textbf{CQELS} & Data-driven & IStream\\
% \hline
% \end{tabular*}
% \vspace*{2mm}
% \caption{Summary table of engine mechanism.}
% \label{tab:RSP-mechanism}
% }
% \end{table}

\subsection{RSP benchmarks}
\textbf{SRBench}, one of the first available RSP benchmarks, proposes a baseline to evaluate the support of various functionalities (SPARQL features, window operator, etc.). \textbf{CSRBench} is an extension of SRBench to  evaluate the results correctness. 
%SRBench and CSRBench are not in the scope of this work, since they do not involve any performance evaluation.

\textbf{LSBench} covers functionality, correctness and performance evaluation. It uses a customized data generator and provides insights into some performance aspects of RSP engines. However, there is no consideration of important performance metrics such as stream rate, window size and number of streams. Besides, the memory consumption has not been considered in their experiments.

\textbf{CityBench} is a recent RSP benchmark based on smart city data and real application scenarios. It provides a consistent and relevant plan to evaluate performance. However, few factors have been considered in the experiments. Only the number of concurrent queries and the number of streams have been considered to evaluate the execution time and memory consumption, whereas other important factors such as window size and stream rate are missing. Moreover, the memory consumption of C-SPARQL shown in CityBench seems to be questionable, since we obtain different results.

% In this paper, we aim to extend existing benchmarks by introducing important factos that have not been considered yet. The evaluation is focused on the two popular engines C-SPARQL and CQELS.
Note that we do not propose a new benchmark for RSP engines. The main goal of this work is to deeply understand the performance of the above-mentioned RSP engines.

\section{Evaluation Plan} \label{sec:evaluation-plan}
\subsection{Dataset and Queries design} \label{Queries}

\begin{figure}[h]
\begin{center}
\includegraphics[width=0.7\linewidth]{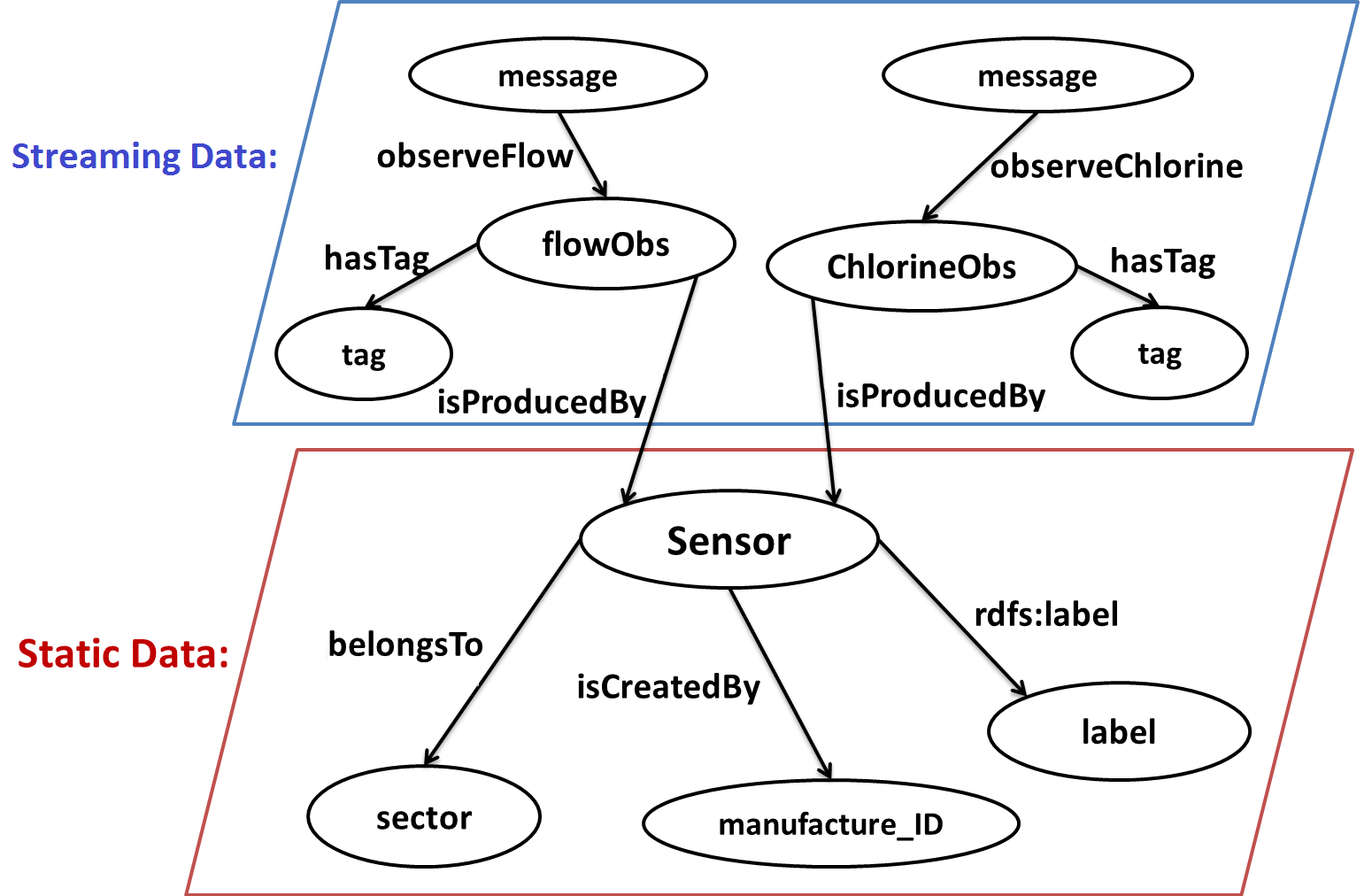}
\captionof{figure}{Dynamic and static data in Water Resource Management Context}
\label{fig:ontology}
\end{center}
\end{figure}

In our experiments, we resolved to use our own data generator for two main reasons: first, to be able to control the size of the generated data streams and, second, to control the data content in order to check the results correctness. In particular, we use both streaming and static data related to the domain of water resource management. The logical data model is presented in Figure \ref{fig:ontology}. The dynamic data describes sensors observations and their metadata, e.g., the message, the observation and the assigned tags. A message basically contains an observation, and we set a fixed number of tags (hasTag predicate) for each observation. For each fifty \emph{flow} observations, we include a \emph{chlorine} observation. The static data provides detailed information about each sensor, namely the label, the manufacturer ID, and the sector ID to which it belongs to in the network. 

%\vspace{-5mm}
%\vspace{-2mm}

We define a set of queries $Q=\{Q_1,Q_2,Q_3,Q_4,Q_5,Q_6\}$, where $Q_1,...,Q_5$ operate over streaming data, and $Q_6$ integrates static data. These queries involve different SPARQL operators (e.g., FILTER, UNION, etc.) and are sorted in ascending order based on the execution complexity (e.g. complex queries involve more query operators). Only the time-based window is addressed in all these queries. As for the last query $Q_6$, we compare the behavior of RSP engines when varying the size of static data. Details and pseudo code of the predefined queries are available on Github\footnote{\url{https://github.com/renxiangnan/Reference_Stream_Reasoning_2016/wiki}}. They can be summarized as follows:
\begin{itemize}
\item $Q_1$: Which observation involves chlorine value?
\item $Q_2$: How many tags are assigned to each chlorine observation?
\item $Q_3$: Which observation ID has an identification ending with ``00" or ``50"?
\item $Q_4$: Which chlorine observation possesses three tags?
\item $Q_5$: Which observation has an identification that ends with ``00" or ``10" and how many tags assigned to this observation?
\item $Q_6$: What is the belonging sector, manufacturer, and assigned label of each chlorine observation?
\end{itemize}

\subsection{Definition of test criteria}
\label{testCriteria}
Let us denote the input parameters by \textbf{$X$}=\{stream rate, number of triples, window size, number of streams, static data size\}, and the set of output metrics by \textbf{$Y$}=\{execution time, memory consumption\}. We next detail each of these parameters.

\textbf{- $X$: (1) stream~rate}~~The time-driven mechanism consists in executing periodically the query with a frequency step specified in the query. This frequency, called \texttt{STEP}, can be time-based (e.g., every 10 seconds) or tuple-based (e.g., every 10 triples). The query is periodically performed over the most recent items. The keyword \texttt{RANGE} defines the size of these temporary items. Just like the frequency step, the window size can be time or tuple-based. In case of time-based window, the execution time and memory consumption are closely dependent on stream rate. Increasing stream rate makes the engines, such as C-SPARQL, process more data for each execution. The frequency step indicates the interval between two successive executions of the same query. Therefore, input stream rate should not exceed engine's processing capacity, otherwise the system has to store an always growing amount of data. 

\textbf{- $X$: (2) number of triples}~~The stream~rate is not an appropriate factor to be considered for the data-driven mechanism because the query execution and the data injection are performed in parallel. In another words, it is not feasible to precisely control the input stream rate. In this context, we need to once feed the system with a fixed number of triples, and that is why we define an additional parameter called \emph{number of triples} $N$. A bigger $N$ generates a smaller error rate, but $N$ should remain under a given threshold to respect the processing limitations of the RSP engines. 

\textbf{- $X$: (3) window size}~~We use \emph{window size} as a performance metric for RSP engines. Note that the window size (\texttt{RANGE}) is closely related to the volume of the queried triples for each execution of the query. According to our preliminary experiments, the window size has marginal impact on the performance of CQELS. Thus, we do not consider this metric when evaluatong CQELS.

\textbf{- $X$: (4) number of streams, (5) static data size}~~The capacity to handle complex queries with multi-stream sources or static background information is an important criterion to evaluate RSP engines. LSBench and CityBench have already proposed these metrics. 
%Note that C-SPARQL supports only one window per query, whereas CQELS defines a window operator on each stream.

\textbf{- $Y$: execution time, memory consumption}~~As the machine conditions are uncontrollable varying factors, we evaluate the execution time, for a given query, as the average value of $n$ iterations. Since C-SPARQL and CQELS have two different execution mechanisms (time-driven and data-driven), we adapt the definition of execution time to each context. In consequence, the execution time represents for C-SPARQL the average execution time over several query executions, while it represents for CQELS the global query execution time for processing $N$ triples.

% The performed experiments are summarized in Table~\ref{tab:criteria}:
% \begin{center}
% \vspace{-1mm}
% \small
% \begin{tabular}{|c|c|c|c|}
% \hline 
% \textbf{RSP Engine} & \textbf{Trigger} & $X$ & $Y$  \\ 
% \hline
% \textbf{C-SPARQL} & Time-driven & \tabincell{c} {Stream Rate, Window Size,\\ Number of Streams, Static Data Size} &\tabincell{c}{Execution Time,\\ Memory Consumption}  \\ 
% \hline
% \textbf{CQELS} & Data-driven & \tabincell{c} {Number of Triples, Number of Streams,\\ Static Data Size} & \tabincell{c}{Execution Time,\\ Memory Consumption}  \\ 
% \hline
% \end{tabular}
% \captionof{table}{Summary table of performance metrics used for each RSP engine.}  
% \label{tab:criteria}
% \end{center}

\section{Experiments} \label{sec:experiment}
All experiments are performed on a laptop equipped with Intel Core i5 quad-core processor (2.70 GHz), 8GB RAM, the maximum heap size is set to 2 GB, running Windows 7, Java version JDK/JRE 1.8. The formal evaluation is done after a 1-to-2-minutes warm-up period with relatively low stream rate.

\subsection{Time-driven: C-SPARQL}
\label{Time-driven: C-SPARQL}
We conducted our experiments over C-SPARQL by testing the previously defined queries. We measure the average value of twenty iterations for query execution time and memory consumption.

% We start our evaluation by measuring the query execution time, then we give some results for memory consumption.

%\vspace{-6mm}
\begin{figure}[h]	\captionsetup{justification=centering,margin=0cm}
	\subfloat[\label{subfig-1}]{%
     	\includegraphics[keepaspectratio=true,scale=0.17]{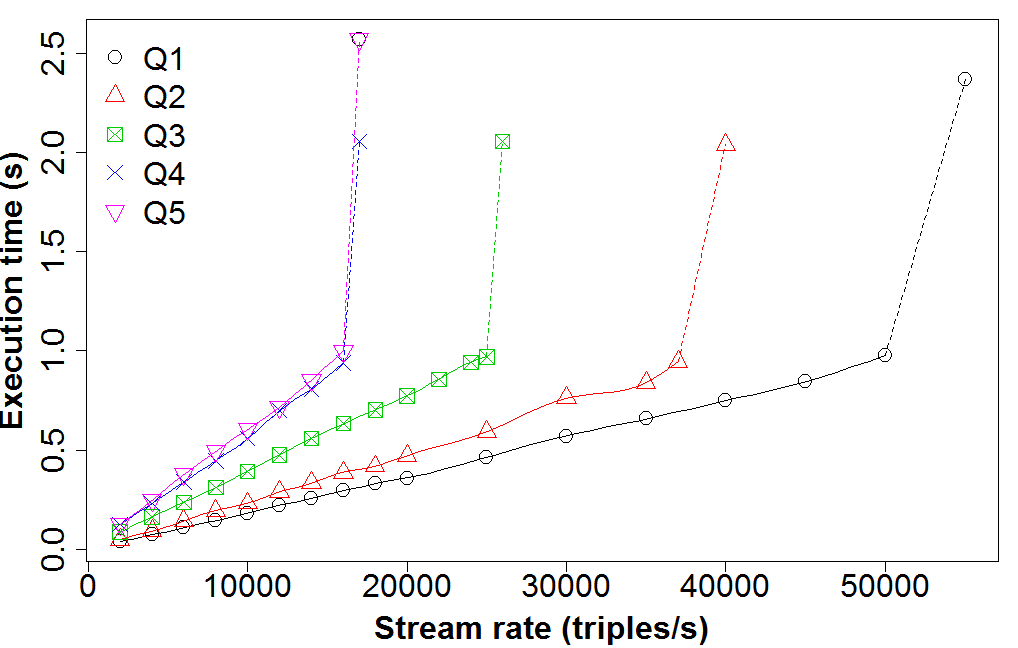}
    }
%    \hfill
   \subfloat[\label{subfig-2}]{%
     	\includegraphics[keepaspectratio=true,scale=0.17]{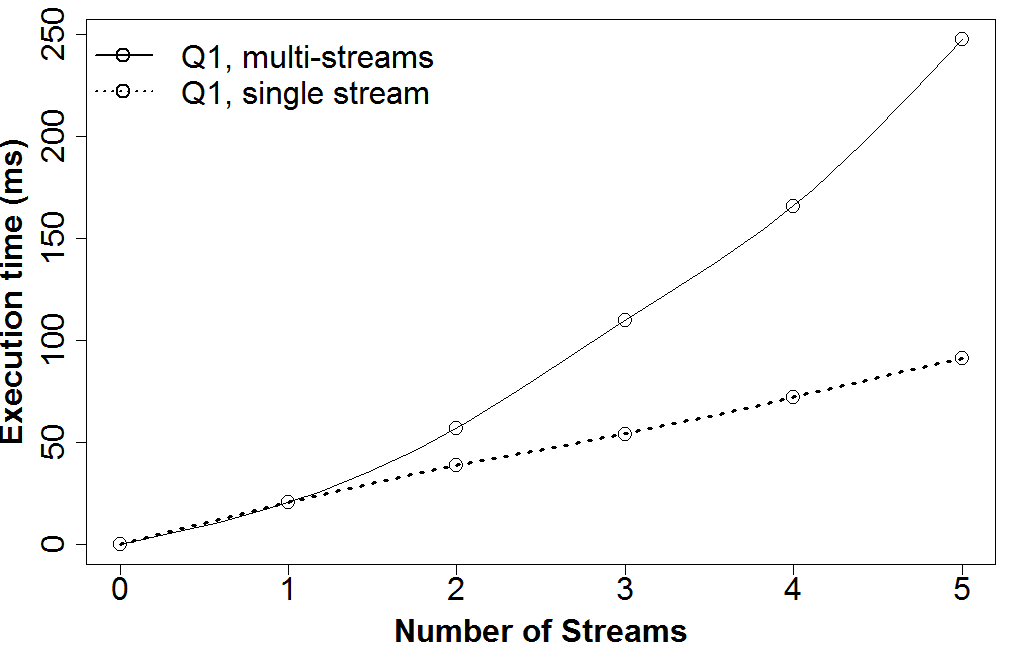}
    }
	\captionof{figure}{Impact of stream rate and number of streams on the execution time of C-SPARQL.}
	\label{fig:C-SPARQL-SR-NS}
\end{figure}
%\vspace{-5mm}

\textbf{Execution Time}
We evaluate query execution time by varying stream rate, number of streams, window size (time-based) and static data size. 

In Figure~\ref{subfig-1}, one can see that the five curves exhibit approximately a linear trend (up to a given threshold concerning the stream rate). %However, it takes longer to process a complex query than a simple query at a given stream rate. 
For each query, the linear trend can be maintained only when the stream rate is under a given threshold. For all five queries, C-SPARQL normally operates when its execution time is smaller than one second, which is also the query preset \texttt{STEP} value. Let us denote by $Rate_{max} (triples/s)$ the maximum stream rate that can be accepted by C-SPARQL for a given query. $Rate_{max}$ represents the maximum number of triples that can be processed per unit time. Table~\ref{tab:Rate_max} shows the $Rate_{max}$ for each query.
% Theoretically, execution time equals to \texttt{STEP} for an incoming stream at $Rate = Rate_{max}$. We can approximate the value of $Rate_{max}$ for each query.
\begin{center}
\begin{tabular}{|c|c|c|c|c|c|}
\hline 
Query & $Q_1$ & $Q_2$ & $Q_3$ & $Q_4$ & $Q_5$ \\ 
\hline
$Rate_{max}$ (triples/s) & $\approx 55000$ & $\approx 40000$ & $\approx 25000$ & $\approx 16000^{{}+{}}$ & $\approx 16000$ \\ 
\hline 
\end{tabular}
\captionof{table}{$Rate_{max}$ for the considered queries in C-SPARQL.} 
\label{tab:Rate_max}
\end{center} 

As shown in Figure~\ref{subfig-1}, if the stream rate exceeds the corresponding $Rate_{max}$, the results provided by C-SPARQL are erroneous. The reason behind is that C-SPARQL does not have enough time to process both current and incoming data. Indeed, newly incoming data stream are jammed in memory, and the system will enforce C-SPARQL to start the next execution which causes errors. Thus, $Rate_{max}$ represents the maximum number of triples under which C-SPARQL delivers correct results.

In some cases, queries require data from multiple streams. In Figure~\ref{subfig-2}, we focus on C-SPARQL's behavior by varying the \textbf{number of streams} where the stream rate is set to 1000 triples/s (i.e. the dotted line in Figure~\ref{subfig-2}). This figure reports the execution time of $Q_1$ for different number of streams. The dotted line represents the execution time of $Q_1$ on a single equivalent (i.e. same workload) stream with a rate $Stream~Rate_{single} = Number~of~Streams \times Stream~Rate_{multi}$, where $Stream~Rate_{single}$ and $Stream~Rate_{multi}$ denote the stream rate for respectively single and multi streams. The curve of query execution time increases as a convex function over the number of streams. C-SPARQL has a substantial delay by the increasing number of streams. Indeed, it has to repeat the query execution for each stream~\cite{CSPARQL1}, then executes the join operation among the intermediate results from different stream sources. This action requires important computing resources, so we can deduce that C-SPARQL is more efficient to process single stream than multi-streams. In addition, according to our experiments, we find that the query execution time linearly increases with the growth of the size of \textbf{time-based window} and \textbf{static data}. C-SPARQL has a constant overhead for delay when increasing these two metrics.

% In Figure~\ref{subfig-2}, we present the impact of time-based \textbf{Window Size} on execution time. Here, we fix stream rate at 1000 triples/s while increasing window size from 2s to 10s using the \texttt{STEP} clause. Figure~\ref{subfig-2} shows that there is a linear relation between window size and execution time for the five queries.

% Query $Q_6$ is specially designed for testing \textbf{Static Data}. We recorded the query execution time by varying the size of static data from 10MB to 50MB. Stream rate is fixed at 1000 triples/s. The curve displayed in Figure~\ref{subfig-4} illustrates the linear increase trend of execution time over static data size. This means that C-SPARQL holds a stable performance while varying the size of background data. The execution time of $Q_6$ is close to one second when 50 MB static data were added. 50 MB is approximately the largest size of static data which can be handled by C-SPARQL for $Q_6$. Experimentation with over 50 MB static data injection emphasizes that C-SPARQL then spends more than one second (i.e. \texttt{STEP} value) to finish its current execution. This will make C-SPARQL unreliable, as the correctness of output will seriously drop.

\textbf{Memory Consumption} We used VisualVM to monitor the Memory Consumption of C-SPARQL. An example of visualization about Java Virtual Machine Garbage Collector (GC) activity is available on our GitHub\footnote{\url{https://github.com/renxiangnan/Reference_Stream_Reasoning_2016/wiki}}.
% \textbf{Memory consumption} on C-SPARQL can be regarded as the activity of GC. One can notice a regular and periodical GC activity during our evaluation on C-SPARQL, i.e., outdated and useless data will be cleaned promptly by GC.

% \begin{center}
% \small
% \begin{tabular}{|c|c|c|c|c|}
% \hline 
% Query & Rate(t/s)  & Period Width (s) & Valley (MB) & Peak (MB) \\ 
% \hline
% $Q_1$ & 100  & 27.40 & 41.14 & 81.07   \\ 
% \hline
% % $Q_2$ & 100  & 51,33 & 42.34 & 124.72  \\ 
% % \hline
% $Q_3$ & 100  & 27,40 & 43.27 & 125.25    \\ 
% \hline
% % $Q_4$ & 100  & 19,29 & 44.38 & 126.72  \\ 
% % \hline
% $Q_5$ & 100  & 23,67 & 44.53 & 148.18   \\ 
% \hline
% $Q_1$ & 1000  & 11.00 & 49.34 & 178.87   \\ 
% \hline
% % $Q_2$ & 1000  & 11.44 & 52.69 & 227.79   \\ 
% % \hline
% $Q_3$ & 1000  & 7.45 & 56.34 & 238.33   \\ 
% \hline
% % $Q_4$ & 1000  & 4.44 & 64.90 & 208.91  \\ 
% % \hline
% $Q_5$ & 1000  & 5.13 & 72.51 & 285.58   \\ 
% \hline
% $Q_1$ & 10000  & 5.00 & 114.28 & 620.01   \\ 
% \hline
% % $Q_2$ & 10000  & 4.00 & 108.52 & 593.74  \\ 
% % \hline
% $Q_3$ & 10000  & 2,27 & 201.73 & 568.79  \\ 
% \hline
% % $Q_4$ & 10000  & 2,86 & 167.67 & 583.81   \\ 
% % \hline
% $Q_5$ & 10000  & 3,12  & 168.91 & 614.82   \\ 
% \hline
% \end{tabular}
% \captionof{table}{The impact of stream rate on the valley value, peak value, and period width of memory consumption.} 
% \label{table:by:fig} 
% \end{center}

%Table~\ref{table:by:fig} shows the evolution of the memory consumption by varying \textbf{Stream Rate} from 100 triples/s to 10000 triples/s. 

\begin{figure}
\centering
\subfloat[\label{subfig-3}]{\label{main:a}\includegraphics[scale=.41]{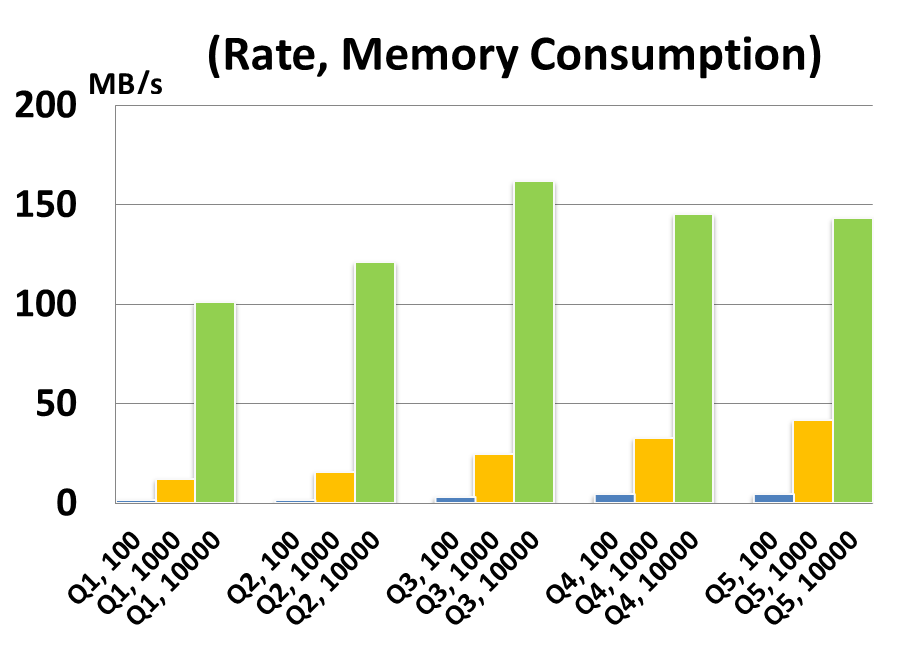}}
\subfloat[\label{subfig-4}]{\label{main:b}\includegraphics[scale=.37]{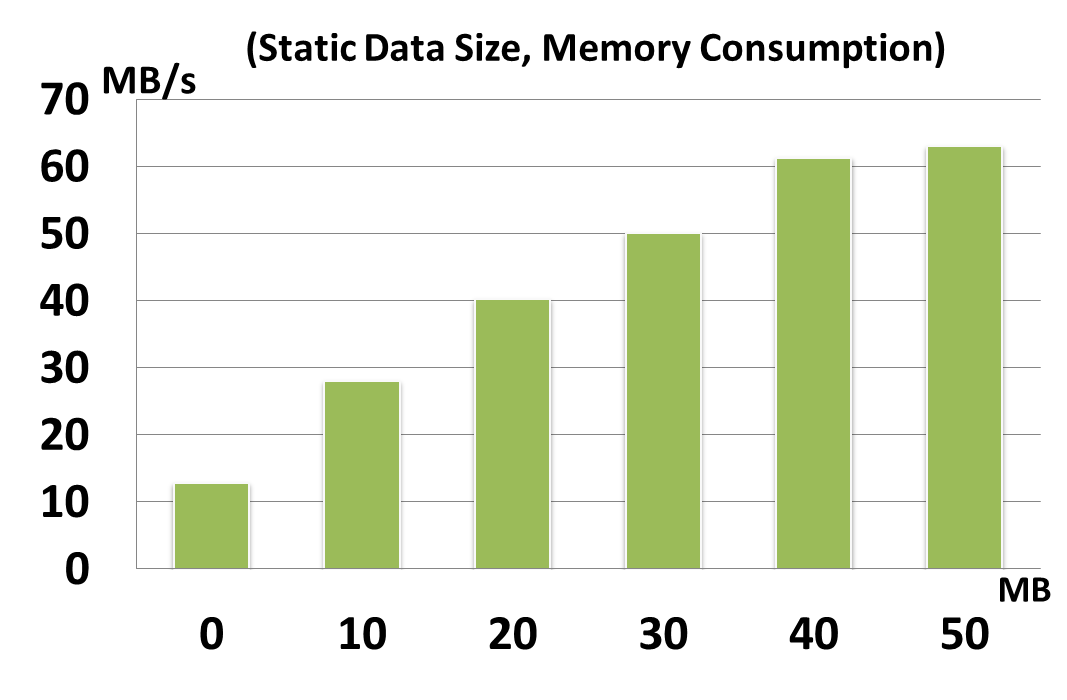}}
\captionof{figure}{The impact of (a) stream rate and (b) static data size on memory consumption in C-SPARQL.}
\label{fig:main}
\end{figure}

Since the Java Virtual Machine executes the Garbage Collector lazily (in order to leave the maximum available CPU resource to the application),
using the maximum memory allocated during execution is not an appropriate way to measure the memory consumption. Practically, the processing of a simple query, while allocating far less memory on each execution, can also reach the maximum allocated heap as the processing of a complex query. Thus, instead, we define a new evaluation metric called Memory Consumption Rate (MCR). Measuring the amount (megabytes) of allocated and freed memory by GC per unit time comprehensively describe MCR. $MCR$(MB/s) $= \frac{\overline{Max}-\overline{Min}}{\overline{Period}}$, $\overline{Max}$ and $\overline{Min}$ refer to the average maximum and minimum memory consumption, respectively. $\overline{Period}$ is the average duration of two consecutive maximum memory observed instances. $\overline{Max}$, $\overline{Min}$, $\overline{Period}$ are computed over 10 observed periods. ${MCR}$ signifies the memory changes in heap per second. A higher $MCR$ shows a more frequent activity of garbage collector. It intuitively shows how many bytes have been released and reallocated by GC per unit time. Figure~\ref{subfig-3} shows the impact of stream rate on $MCR$. For each query, the period decreases and $MCR$ increases with the growth of Stream Rate. Query $Q_3$ has the highest $MCR$. This can be explained by the aggregate operator which produces more intermediate results during query execution. Note that $MCR$ is not a general criterion for measuring memory consumption. In some use cases, we could not observe periodical activity on GC. The main goal of using $MCR$ is to give a comprehensive description of memory management on C-SPARQL.

%\vspace{-7mm}
%\vspace{-5mm}

Figure~\ref{subfig-4} displays the increase of memory consumption rate over the growth of \textbf{static data size}. For query $Q_6$, memory peak varies marginally while increasing static data size, but the minimum consumed memory is directly impacted. One possible explanation is that C-SPARQL produces additional objects to process static data, and keeps these objects as long-term in memory. 

\subsection{Data-driven: CQELS}
\label{Data-driven: CQELS}
This section focuses on the performance evaluation of CQELS. The variant parameters are number of triples, number of streams, and static data size. $Q_4$ and $Q_5$ are not included in this evaluation since CQELS does not support the timestamp function (i.e. function that performs basic temporal filtering on the streamed triples).

\textbf{Execution Time} Since CQELS uses a so-called probing sequence to support its execution plan, 
% \textcolor{red}{There's no way to distinguish whether a mapping is a final result or an intermediate result, and the system will only notify if a final result is produced. In other words, getting the execution time for each triple is not experimentally feasible.}
getting the running time for each query execution is not experimentally feasible. Thus, we evaluate the global execution time of $N$ triples for CQELS. More precisely, we keep the same strategy as LSBench, i.e. inject a finite sequence of stream into the system which contains $N$ triples. $N$ should be big enough to get more accurate results ($N \ge 10^5$).   

\begin{figure}[h]
\vspace{-5mm}
\captionsetup{justification=centering,margin=0cm}
\subfloat[\label{subfig-5}]{%
\includegraphics[keepaspectratio=true,scale=0.17]{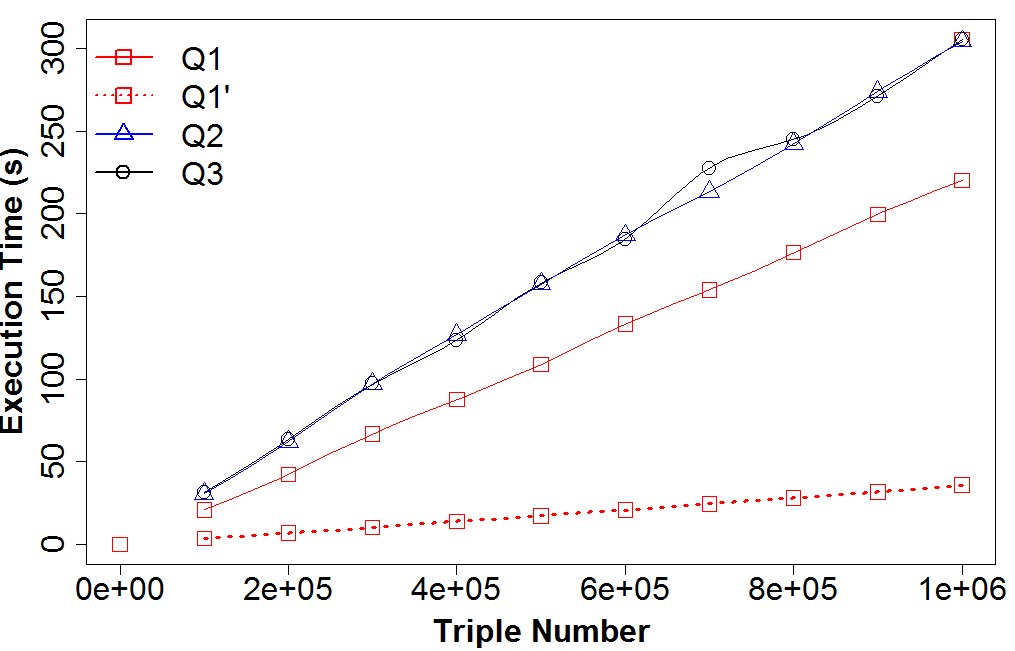}}
\subfloat[\label{subfig-6}]{%
\includegraphics[keepaspectratio=true,scale=0.17]{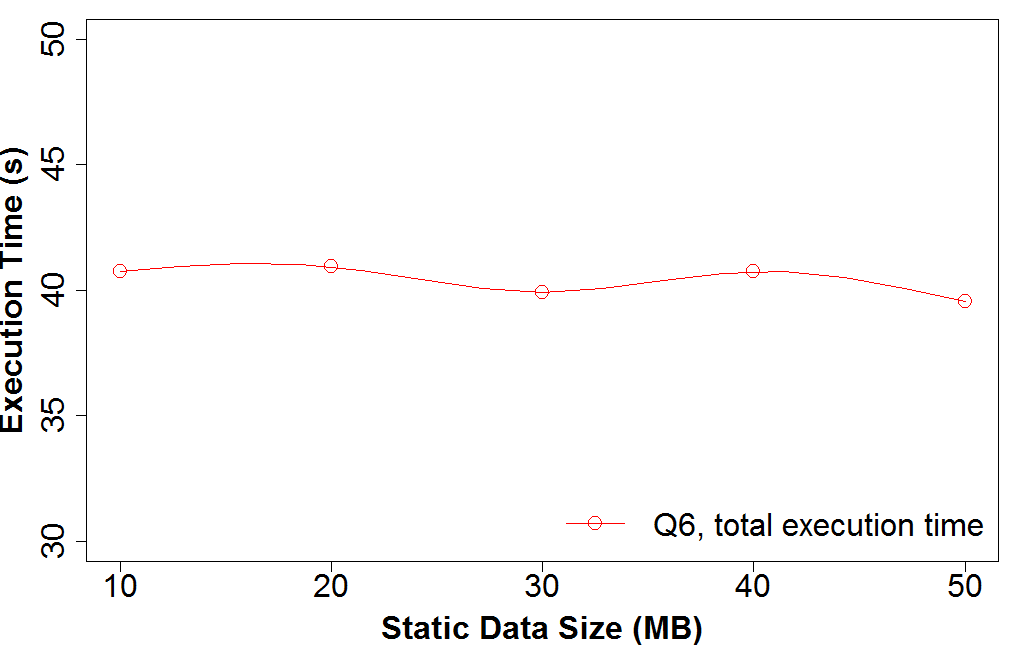}}
\captionof{figure}{The impact of number of triples and static data size on query execution time in CQELS. }
\label{Asy_Streams_Static_Data}
\vspace{-5mm}
\end{figure}

Figure~\ref{subfig-5} shows the impact of \textbf{number of triples} on execution time. $N$ should also be controlled within a certain range to prevent the engine from crashing (c.f. ``Memory Consumption" part of CQELS). Queries $Q_1$, $Q_2$, $Q_3$ contain chain patterns (join occurs on subject-object position) that select chlorine observation: \emph{\{$T_1$: ?observation ex:observeChlorine ?chlorineObs .} $T_2$: \emph{?chlorineObs ex:hasTag ?tag .  \}}. Pattern $T_1$ returns all results by matching the predicate ``observeChlorine", then $T_2$ filters among all selected observations in $T_1$ those which have been assigned tags. In Figure~\ref{subfig-5}, note that there is no significant difference between $Q_2$ and $Q_3$. Based on $Q_2$, the query $Q_3$ adds a ``FILTER" operator to restrict that preselected observations which have an ID ending by ``00" or ``50". This additional filter in $Q_3$ slightly influences the engine performance, which lets suggest that CQELS is very efficient at processing ``FILTER" operator. As the dotted line $Q_1'$, it represents $Q_1$ without the pattern $T_2$. Its corresponding execution time is reduced to one-six times compared with $Q_1$. Indeed, the pattern $T_2$ plays a key role in term of execution time. Without $T_2$, CQELS will return the results immediately if $T_1$ is verified, but pattern $T_2$ makes the engine wait till $T_2$ is verified.  

CQELS supports queries with \textbf{multi-streams}. It allows to assign the triple patterns which are only relative to the corresponding stream source. This property gives the engine some advantages to process complex queries. Each triple just needs to be verified in its attached stream source. However, C-SPARQL has to repeat verification on all presenting streams for the whole query syntax, and this behavior leads to a waste of computing resources. Due to data-driven mechanism, serious mismatches occur in output for a multi-streams query, especially when the query requests synchronization among the triples. Asynchronous streams are illustrated in our GitHub\footnote{\url{https://github.com/renxiangnan/Reference_Stream_Reasoning_2016/wiki}}.

Suppose that we have two streams, $S_1$ and $S_2$, sequentially sent (due to the data-driven approach adopted by CQELS) into the engine. If the window size defined on $S_1$ is not large enough, $?observation$ in pattern $T_2$ will not be matched with $?observation$ in $T_1$. This problem can be solved by defining a larger window size in $T_1$ with a small number of streams. In our experiments, we carry out the multi-streams test by constructing two streams on $Q_1$, $Q_2$ and $Q_3$. For $Q_1$, with two streams, CQELS spent approximately 26s to process $(2 \times)~10^5$ triples, that is just 30\% more than the single stream case. To conclude, CQELS gains some advantages in term of execution time to process queries with multi-streams. However, the output may also be influenced by the asynchronous behavior in multi-stream context. Note that C-SPARQL does not suffer from the streams synchronization since it follows batch-oriented approach.

In Figure~\ref{subfig-6}, the curve gives the total execution time(s) for 1.260.000 triples. The execution time for $N$ triples slightly changes while increasing the size of \textbf{Static Data} from 10MB to 50MB. The result shows that CQELS is efficient for processing static data of a large size.

\textbf{Memory Consumption} As we directly send $N$ triples into the system at once, CQELS's memory consumption does not behave as C-SPARQL (which follows a periodic pattern). Generally, the memory consumption on CQELS keeps growing by increasing the number $N$ of triples. As mentioned in the previous section, $N$ should not exceed a given threshold. If $N$ is very large, the memory consumption will reach its limit. In this situation, latency on query execution will increase substantially. Furthermore, since serious mismatch occurs on multi-streams query, $X$ = Number of Stream is not considered as a metric for memory consumption. We evaluate the peak of memory consumption (MC) during query execution. The trend increases over time, where $MC$ reaches the peak just before the end of query execution.  

\begin{center}
\begin{figure}[h]
	\captionsetup{justification=centering,margin=0cm}
	\subfloat[\label{subfig-9}]{%
     	\includegraphics[keepaspectratio=true,scale=0.3]{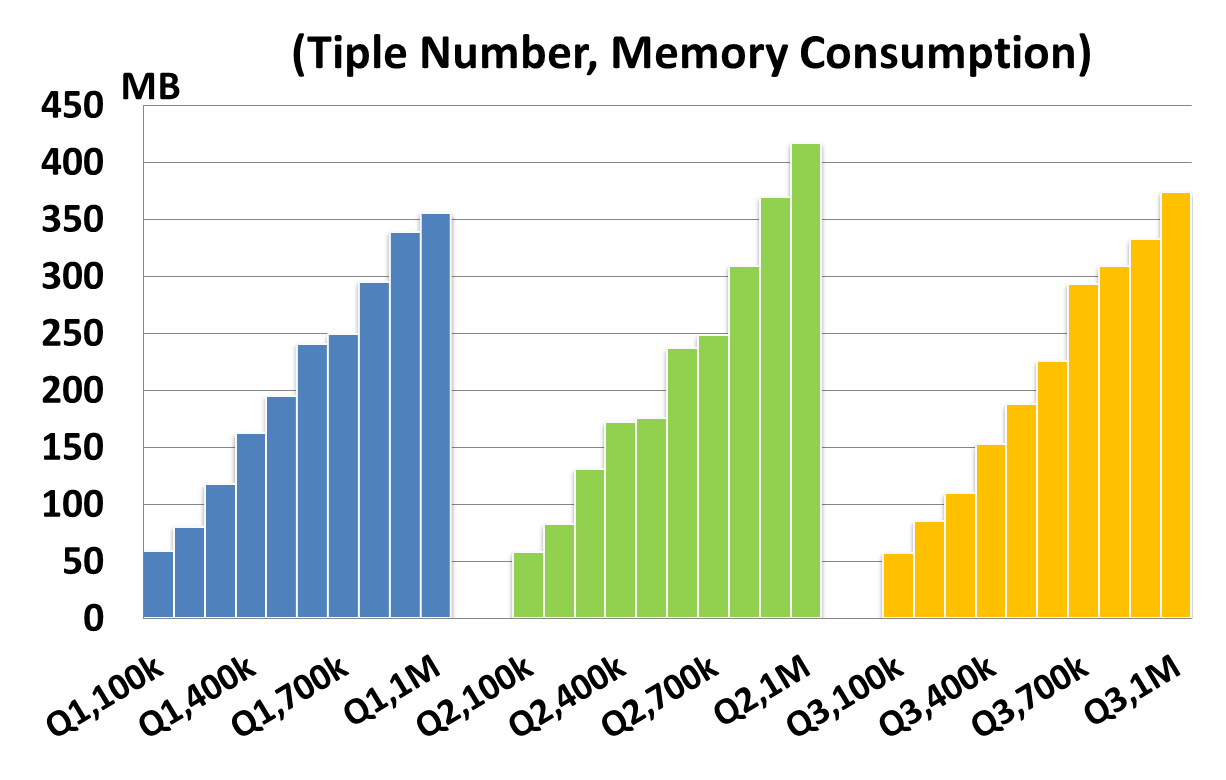}
    }
   \subfloat[\label{subfig-10}]{%
     	\includegraphics[keepaspectratio=true,scale=0.3]{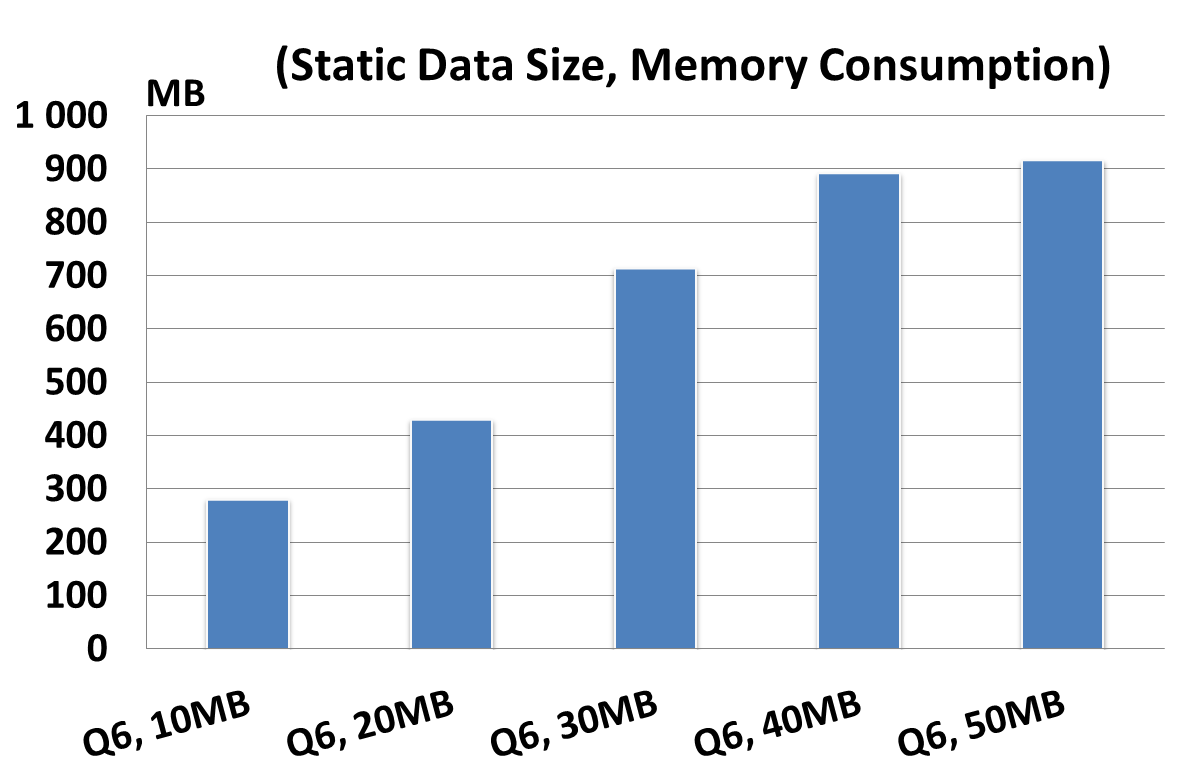}
    }
	\captionof{figure}{Impact of the number of triples and the static data size on memory consumption in CQELS. }
	\label{MC_CQELS}
\end{figure}
\end{center}

Figure~\ref{subfig-9} shows that the memory consumption of $Q_1$, $Q_2$ and $Q_3$ is very close when varying the \textbf{number of triples}, i.e., the complexity of queries are not reflected by their memory consumption. CQELS manages efficiently the memory for complex queries. In Figure~\ref{subfig-10}, the memory consumption of $Q_6$ is proportional to the size of \textbf{static data}. According to the evaluation, we found that a lower maximum allocated heap size (e.g., 512MB) causes a substantial delay on CQELS. The consumed memory keeps growing to the limited heap size, i.e. the GC could clear the unused objects in a timely manner. This behavior is possibly caused by the built-dictionary for URI encoding \cite{CQELS}.

\vspace{-5mm}
\vspace{-10mm}

\section{Results Discussion} \label{sec:result-discussion}
\label{Discussion}

As we generate different streaming modes for time-driven (C-SPARQL) and data-driven (CQELS) engines, the memory consumption is not comparable between them. This section mainly derives a discussion on query execution time based on observed results. It is about a simple comparison between C-SPARQL and CQELS.

It is not obvious to compare the performance of different RSP engines, since each of them has a specific execution strategy. According to \cite{LSBench} and to our experiments, we list the following conditions to support a fair cross-engines performance comparison: (i) the engine results should be correct, at least comparable. We remind that the untypical behavior of C-SPARQL occurs when the incoming stream rate exceeds the threshold. Even if the engine still produces results, it is meaningless to measure the execution time; (ii) The execution time for different RSP engines should associate the same workload. As C-SPARQL uses a batch mechanism, it is easy to control the workload of the window operator. However, the data-driven eager mechanism practically makes infeasible the workload control. Therefore, we choose $t = \frac{T}{N}$, the average execution time per triple to support our comparison. $T$ is the total execution time for $N$ triples. Note that $t$ marginally changes when varying the metrics defined in section \ref{testCriteria}; (iii) The engine warming up is also recommended. We inject the ``warming up" stream (with a relatively low stream rate) into the system before the formal evaluation.

% \begin{center}
% \begin{tabular}{|c|c|c|c|c|c|c|}
% \hline 
% &\multicolumn{6}{c|}{Execution Time (second)} \\ 
% \hline 
% RSP engines& $Q_1$ & $Q_2$ & $Q_3$ & $Q_4$ & $Q_5$ & $Q_6$ \\ 
% \hline
% CSPARQL & 21.70 & 26.61 & 49.43 & 70.37 & 76.26 &1100.21 \\ 
% \hline 
% CQELS & 213.21 & 300.75 & 305.636 & - & - &39.86  \\
% \hline  
% \end{tabular}
% \captionof{table}{Execution time of $Q_1$, $Q_2$, $Q_3$, $Q_4$, $Q_5$ and $Q_6$} 
% \label{tab:comparison}
% \end{center} 

\begin{center}
\begin{tabular}{|c|c|c|c|c|}
\hline 
&\multicolumn{4}{c|}{Average execution time per triple (millisecond)} \\ 
\hline 
RSP engine & $Q_1$ & $Q_2$ & $Q_3$ & ~~$Q_6$ (50MB static data)~~ \\ 
\hline
CSPARQL & ~~0.018~~ & ~~0.025~~ & ~~0.040~~ & 0.952 \\ 
\hline 
CQELS & 0.169 & 0.239 & 0.243 & 0.032  \\
\hline  
\end{tabular}
\captionof{table}{Execution time (in seconds) of $Q_1$, $Q_2$, $Q_3$ and $Q_6$.} 
\label{tab:comparison}
\end{center}

Table \ref{tab:comparison} shows that C-SPARQL outperforms CQELS to deal with $Q_1$, $Q_2$ and $Q_3$. This can be explained by the fact that the chain pattern existing in $Q_1$, $Q_2$ and $Q_3$ forces CQELS to repeat the verification on matching condition for the whole window. This behavior significantly hinder the engine performance. For $Q_6$, CQELS is almost 27 times faster than C-SPARQL. It shows its high efficiency to process queries with static data. 

Finally, we summarize our experiment over three aspects: 1) \textbf{Functionality support}. Since C-SPARQL uses the Sesame/Jena as the querying core, it supports most of the SPARQL 1.1 grammar. In contrast, as CQELS is implemented in a native way, it supports less operations than C-SPARQL, e.g., timestamp function, property path, etc. 2) \textbf{Output correctness}. As mentioned in section \ref{Data-driven: CQELS}, CQELS suffers from a serious output mismatch in the multi-stream context. This is due to the eager execution mechanism and asynchronous streams. C-SPARQL behaves normally with multi-stream queries since it is characterized by a time-driven mechanism. As a matter of fact, real use cases often require concurrency of join from different stream sources. In this context, C-SPARQL takes the advantages of correctness and completeness of output results. 3) \textbf{Performance}. C-SPARQL shows stability with complex queries. However, in practical applications, input stream rate should be controlled at a low level to guarantee C-SPARQL's output correctness. Besides, C-SPARQL has scalability problem when dealing with static data. CQELS takes advantage from its dictionary encoding technique and dynamic routing policy, and thus, is efficient for simple queries and is scalable with static data.

\section{Conclusion} 
\label{sec:conclusion}
This paper focuses on the performance evaluation of two state-of the-art engines: C-SPARQL and CQELS. We propose some new performance metrics and designed a specific evaluation plan. In particular, we take into account the specific implementation of each RSP engine. We performed many experiments to evaluate the impact of \emph{Stream Rate}, \emph{Number of Triples}, \emph{Window Size},  \emph{Number of Streams} and \emph{Static Data Size} on \emph{Execution Time} and \emph{Memory Consumption}. Several queries with different complexities have been considered. The main result of this complete study is that each RSP engine has its own advantage and is adapted to a particular context and use case, e.g., C-SPARQL excels on complex and multi-stream queries while CQELS stands out on queries requiring static data. In future work, we plan to evaluate the performance of RSP engines in a distributed environment. 

\section*{Acknowledgments}
This work has been supported by the WAVES project which is partially supported by the French FUI (Fonds Unique Interministériel) call \#17.

\bibliographystyle{abbrv}
\bibliography{SR2016}
\end{document}